# Novel K Rattling: A New Route to Thermoelectric Materials?

Elvis Shoko[1a)], Y. Okamoto[2], Gordon J Kearley[1], Vanessa K Peterson[1], Gordon J Thorogood[1]

1. Australian Nuclear Science and Technology Organisation Locked Bag 2001, Kirrawee DC, NSW 2232, Australia
2. Institute for Solid State Physics, University of Tokyo, Kashiwanoha, Kashiwa, Chiba 277-8581, Japan.

## Abstract

We have performed *ab initio* molecular dynamics (MD) simulations to study the alkali-metal dynamics in the Al-doped ($KAl_{0.33}W_{1.67}O_6$ and $RbAl_{0.33}W_{1.67}O_6$) and undoped ($KW_2O_6$ and $RbW_2O_6$) defect pyrochlore tungstates. The K atoms exhibit novel rattling dynamics in both the doped and undoped tungstates while the Rb atoms do not. The $KAl_{0.33}W_{1.67}O_6$ experimental thermal conductivity curve shows an unusual depression between ~ 50 K and ~ 250 K, coinciding with two crossovers in the K dynamics: the first at ~ 50 K, from oscillatory to diffusive, and the second at ~ 250 K, from diffusive back to oscillatory. We found that the low-temperature crossover is a result of the system transitioning below the activation energy of the diffusive dynamics whereas the high-temperature crossover is driven by a complex reconstruction of the local potential around the K atoms due to the cage dynamics. This leads to a hardening of the K potential with increasing temperature. This unusual reconstruction of the potential may have important implications for the interpretation of finite-temperature dynamics based on zero-temperature potentials in similar materials. The key result is that the novel K rattling, involving local diffusion, leads to a significant reduction in the thermal conductivity. We suggest that this may open a new route in the phonon engineering of cage compounds for thermoelectric materials where the rattlers are specifically selected to reduce the lattice thermal conductivity by the mechanism of local diffusion.

[a)] Author to whom correspondence should be addressed. Electronic mail: elvis.shoko@gmail.com.
Telephone: +61-41-506-4823

# I. Introduction

Thermoelectric power generation holds the prospect of a clean technology for converting waste heat from automobiles and industrial processes to electricity thereby improving their overall efficiencies.[1] To be successful, one of the major hurdles the technology needs to overcome is the low efficiencies of the materials used in the thermoelectric generators. The material efficiency is defined by its thermoelectric figure of merit, ZT:

$$ZT = \frac{S^2 \sigma T}{\kappa_l + \kappa_e} \tag{1}$$

Where $S$ is the Seebeck coefficient, $\sigma$, the electrical conductivity, $T$, the absolute temperature, and $\kappa_l$ and $\kappa_e$ are the lattice and electronic thermal conductivities, respectively. Thermoelectric technology requires thermoelectric materials with ZT ≥ 3 to be commercially viable and competitive with conventional power generation technologies[2]. Thus the goal in materials development is to maximize ZT but this is complicated because increasing $\sigma$ in Eq. (1) also increases $\kappa_e$ through the Wiedemann-Franz law: $\kappa_e = L\sigma T$, where $L$ is the Lorentz number. A widely used strategy to deal with this conflict is to seek compounds which have been termed phonon-glass-electron-crystal (PGEC) materials.[3, 4] The idea of a PGEC material is that its thermal conductivity is glass-like (low $\kappa_l$) while the electrical conductivity is similar to that of a narrow gap crystalline semiconductor. A class of so called cage compounds which includes clathrates,[1, 5, 6] skutterudites,[3, 4, 7, 8] and β-pyrochlores,[9, 10] considered promising candidate PGEC materials are extensively studied for thermoelectric development. In these materials, the phonon glass property is considered to come from weakly bound atoms (rattlers) exhibiting low-energy vibrational modes which scatter the heat-carrying acoustic phonons of the lattice. The precise mechanism by which a rattler scatters phonons is still the subject of research with both resonant scattering[11-13] and localized mode coupling[5, 6, 8, 14, 15] having been proposed. The distinction between the two mechanisms is perhaps easily understood from the simple (kinetic theory of gases) expression for thermal conductivity:

$$\kappa = \frac{1}{3} C v l \tag{2}$$

Where $C$ is the specific heat capacity, $v$, the group velocity, and $l$, the mean free path. In resonant scattering $\kappa$ is reduced by a reduction in $l$ while $v$ is reduced in localized mode coupling. Despite this lack of a clear mechanistic picture of rattler-phonon interaction, experiments show that filling cage compounds with suitable rattler atoms reduces the thermal conductivity.[16] Even further reductions have been reported in multi-filled compounds where two or more elements are used as rattler atoms.[7, 17-24] The larger reductions in the thermal conductivity in the thermal conductivity of multi- compared to single-filled materials is attributed to the scattering of a broader range of host lattice phonon frequencies due to the rattlers of the different elements vibrating at different frequencies. In a recent contribution[25] on the study of rattling in the Al-doped β-pyrochlores, $AAl_{0.33}W_{1.67}O_6$ (A = K, Rb, Cs), we reported the discovery of novel rattling of K atoms in $KAl_{0.33}W_{1.67}O_6$ involving both vibrations at a broad range of frequencies as well as local diffusion. We suggested that the novel K dynamics could be of interest in the development of new thermoelectric materials. In particular we noted that the combination of a broad range of vibrational frequencies and local diffusion in the dynamics of a single rattler could be more effective at scattering lattice phonons compared to a rattler exhibiting only one of these effects. In this work, we examine this conjecture through a study of the thermal conductivities of the Al-doped β-pyrochlores. For these compounds, the A cations are the rattlers and they reside inside large $Al_2W_{10}O_{18}$ octahedral cages where they may occupy either a cage center (8*b*) site or an off-center (32*e*) site as illustrated in FIG. 1.

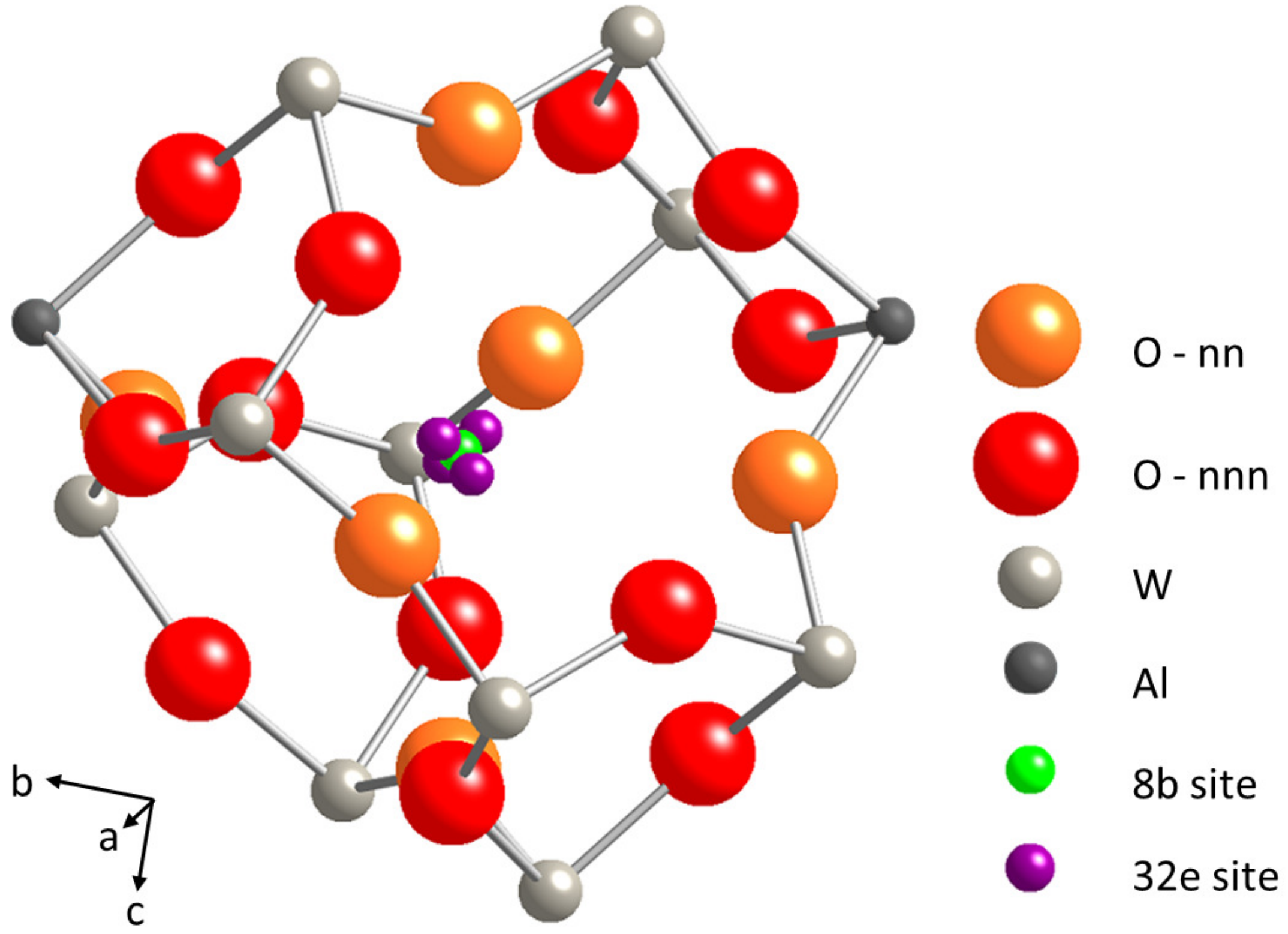

**FIG. 1. Illustration of the cage structure of the Al-doped defect pyrochlore tungstate. The cage consists of six nearest neighbour O atoms (O – nn) octahedrally arranged around the cage center followed by twelve next-nearest neighbour O atoms (O – nnn). There are twelve W lattice sites (16*c*) per cage, two of which are doped with Al. The alkali metal atom can occupy either of two sites inside the cage, the first labeled as 8*b* site at the cage center and the second is the 32*e* site.**

**The four symmetry-equivalent 32*e* sites per cage form a regular tetrahedron centered at the cage centre. The axes are the lattice vectors of the tetragonal supercell used in the MD simulations.**

In the K, Rb, and Cs pyrochlore series, the ratios of the rattler ionic radius to the lattice constant of the respective pyrochlores are: 0.136, 0.149, and 0.164. The increasing ratio shows that the change in the ionic radius from K to Cs is larger than the corresponding change in the lattice constant of the pyrochlore. Consequently, this series affords a study of the effect of the size of the rattler on the rattling dynamics.[26] In this study, we use *ab initio* molecular dynamics (MD) to explain the relatively low thermal conductivity of $KAl_{0.33}W_{1.67}O_6$ and some unusual features in the conductivity curve. We first show further evidence that the diffusive vibrational spectrum of this compound is a consequence of the small size of the K rattler relative to its cage volume. We then discuss the measured thermal conductivities of all three pyrochlores and show that the electronic contributions are negligible for all three, so that any differences reflect differences in their lattice thermal conductivities. The $KAl_{0.33}W_{1.67}O_6$ thermal conductivity curve is characterized by an anomalous depression for 25 K < T < 250 K which is a consequence of the rattler undergoing local diffusion in this temperature range. We show that the combination of the highly anharmonic and anisotropic K potentials and entropic effects leads to unusual temperature dependence for the K dynamics involving a cross-over from oscillatory to diffusive dynamics at ~ 25 K < T < 50 K and back to oscillatory dynamics at ~ T > 250 K. We then conclude by suggesting a new approach to the multi-filling method which expands the approach to include rattlers which absorb the lattice phonons through local diffusion. We believe this opens a new and viable approach to phonon engineering in thermoelectric material development.

# II. Experiment and Simulations

Thermal conductivity measurements between 10 and 300 K were performed in a Physical Properties Measurement System with the thermal transport option (Quantum Design).

For the doped tungstates, *ab initio* MD simulations were performed using a *3* x *1* x *1* supercell (216 atoms) of each structure consisting of 8 Al atoms substituting a selection of the 48 W atoms within the cell.[25] As there is no unique way of making this selection, the choice was guided by the likely scenario where the Al atoms are well separated from each other. Two additional MD simulations were performed on the undoped tungstates, $KW_2O_6$ and $RbW_2O_6$, using their respective conventional unit cells (72 atoms). As synthesis of these compounds has not been reported in the literature, there are no published lattice parameters for these tungstates so we used our data for the Al-doped counterparts in these calculations. The Projector-Augmented Wave (PAW) method[27, 28] implemented in the Vienna Ab-initio Simulation Package (VASP)[29, 30] was employed, and for the exchange-correlation potential, the generalized gradient approximation with the Perdew, Burke, and Ernzerhof (GGA-PBE) functional was used.[31, 32]

All structures were first relaxed to obtain optimized geometries within the experimental lattice parameters before equilibration at the desired temperature. Microcanonical ensemble (NVE) runs were then performed with Γ-point sampling for a minimum of 24 ps with a time step of 1 fs. For the electronic structure calculations, we used an 8x8x8 k-point mesh with a 400 eV energy cut off.

# III. Results and Discussion

All analysis of the vibrational dynamics is performed in the principal axis coordinates of the motion of the atoms.[33] In order to discuss the spectral properties of the vibrational dynamics, we calculate magnitude spectra from the (forward) fast Fourier transform (FFT) of the atomic coordinates of the MD trajectories. For the magnitude spectra, we are only interested in comparing spectral forms, and so the magnitudes are suitably scaled to facilitate this type of analysis. Comparing actual absolute values of the magnitudes in the spectra is not meaningful as these depend on a number of variables, e.g., simulation length and temperature, which may vary between different simulations. As a tool for assessing the nature of the diffusive motion, we calculate the mean-square displacement autocorrelation function (msd):[34, 35]

$$msd(j) = \frac{1}{N-j}\sum_{i=1}^{N-j}\left[\mathbf{R}(t_i) - \mathbf{R}(t_{i+j})\right]^2, \quad j = 1,2,...,N-1 \tag{3}$$

Where $\mathbf{R}(t_i)$ and $\mathbf{R}(t_{i+j})$ are the nuclear coordinates at the $t_i$th and $t_{i+j}$th simulation time steps, respectively and $N$ is the total number of time steps in the simulation. In this work, the msd is calculated as an average over all the 24 alkali-metal atoms in the simulation cell for the doped compounds, and 8 for the undoped. The physical meaning of the msd function is understood by considering two simple limiting cases of the dynamics of an atom in a solid. In the first case, the atom undergoes long-range diffusion such that its distance from the starting point is always increasing. In this case, the msd function is a monotonically-increasing function of time. In the second case, the atom vibrates harmonically about its mean position so that its displacement in one dimension (1D) (e.g., along $x$) can be written as: $x(t) = A * \sin(\omega t + \varphi)$, where $A$ is the amplitude, $\omega$ is a characteristic frequency (rads/s), and $\varphi$ is the phase angle. For this case, the msd function oscillates between 0 and $2 * A^2$ with frequency $\omega$. Real dynamics of atoms fall somewhere between these extremes; approaching the monotonically-increasing function limit if they undergo long-range Fickian diffusion, or the uniformly-oscillating function if the local potential is approximately harmonic.

We study the local potentials around the K atoms at different temperatures by calculating the potential of mean force (PMF):[36]

$$PMF = -k_B T * \log P(\mathbf{r}) \tag{4}$$

With $k_B$, the Boltzmann constant, $T$, absolute temperature, and $P(\mathbf{r})$ the population density at position $\mathbf{r}$ for a given atom. In 1D, the PMF is the free energy curve along a specified direction and consists of the potential energy part and the entropy. We calculate a PMF along the direction of maximum displacement for the K atoms as an average over all these atoms in the simulation cell. Each atom is centered at its time-average position and displacements are then calculated relative to this position. The full range of the displacements for all the atoms is then partitioned into suitable intervals from which a histogram of the displacements is constructed for all the atoms over the duration of the simulation. The population density, $P(\mathbf{r})$ in Eq. (4), is calculated from this histogram.

For all the simulations, the system temperatures gradually decreased from their target values as a result of the inherent rounding errors in the numerical algorithms. For each target temperature, we give the actual average value calculated over the duration of the simulation in parenthesis: 25 K (17), 50 K (40), 100 K (60), 150 K (147), 250 (244), 400 K (387), and 1000 K (910) for the $KAl_{0.33}W_{1.67}O_6$ runs. For the 100 K undoped tungstate simulations, the average temperatures were 87 K and 83 K for $KW_2O_6$ and $RbW_2O_6$, respectively.

### *A. Novel Rattling Dynamics of K Atoms*

In our previous work,[25] we showed that the K atoms in $KAl_{0.33}W_{1.67}O_6$ undergo novel dynamics distinct from the K in the undoped osmate analogue, $KOs_2O_6$. Here, we first compare the alkali-metal dynamics between the Al-doped and undoped tungstates. In FIG. 2, we plot the magnitude spectra of K and Rb in $KAl_{0.33}W_{1.67}O_6$ and $RbAl_{0.33}W_{1.67}O_6$, respectively, alongside their undoped analogues to examine if the novel K rattling previously reported is a result of the Al doping. Both K spectra exhibit increasing magnitude with decreasing energy at low energies ($E < 3$ meV), i.e., a quasielastic spectral form, indicative of diffusive dynamics. In contrast, Rb spectra exhibit no rising tail a low energies for both spectra, indicating that these atoms undergo oscillatory dynamics. These results corroborate our previous findings[25] that the rattler size plays a significant role in explaining the novel K dynamics. Figure 2 also shows that there are some differences in the spectra between the doped and undoped counterparts for each rattler. These differences may be attributed to the effects of Al doping on the rattler dynamics, but they do not appear to change the essential rattler dynamics in a significant way.

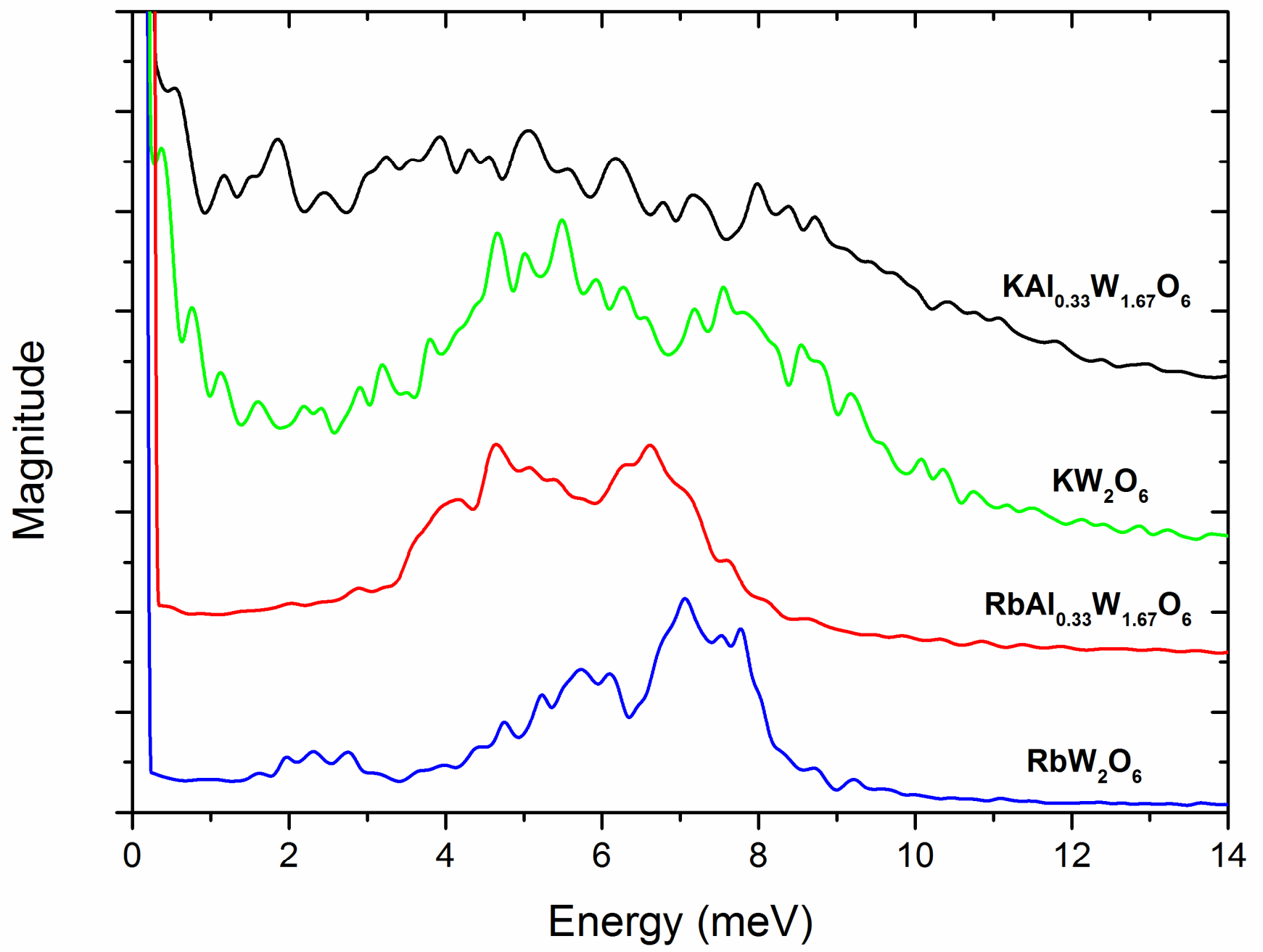

**FIG. 2. Magnitude spectra calculated from MD simulations at 100 K for the K and Rb atoms in the different compounds as indicated. Both K spectra exhibit a quasielastic profile which is absent from the Rb spectra, indicating an essential difference in the dynamics of these atoms. Spectra are suitably scaled and displaced along the magnitude axis for clarity.**

The msd autocorrelation functions plotted in FIG. 3 clarify the nature of the diffusive K dynamics we observe in FIG. 2. Both K systems show distinct profiles for the msd function in one direction only – the direction of maximum displacement. The profiles in the other directions are similar to the Rb systems, with relatively flat msd functions characterized by small oscillations around the mean positions, consistent with oscillatory dynamics. In contrast, the K atoms undergo local diffusion in the direction of maximum displacement as can be judged from the rising and falling of the msd function over longer time intervals compared to the other directions. By tracking a single K atom, we showed, in our previous work,[25] that the K atoms diffuse between two sites in a double well potential.

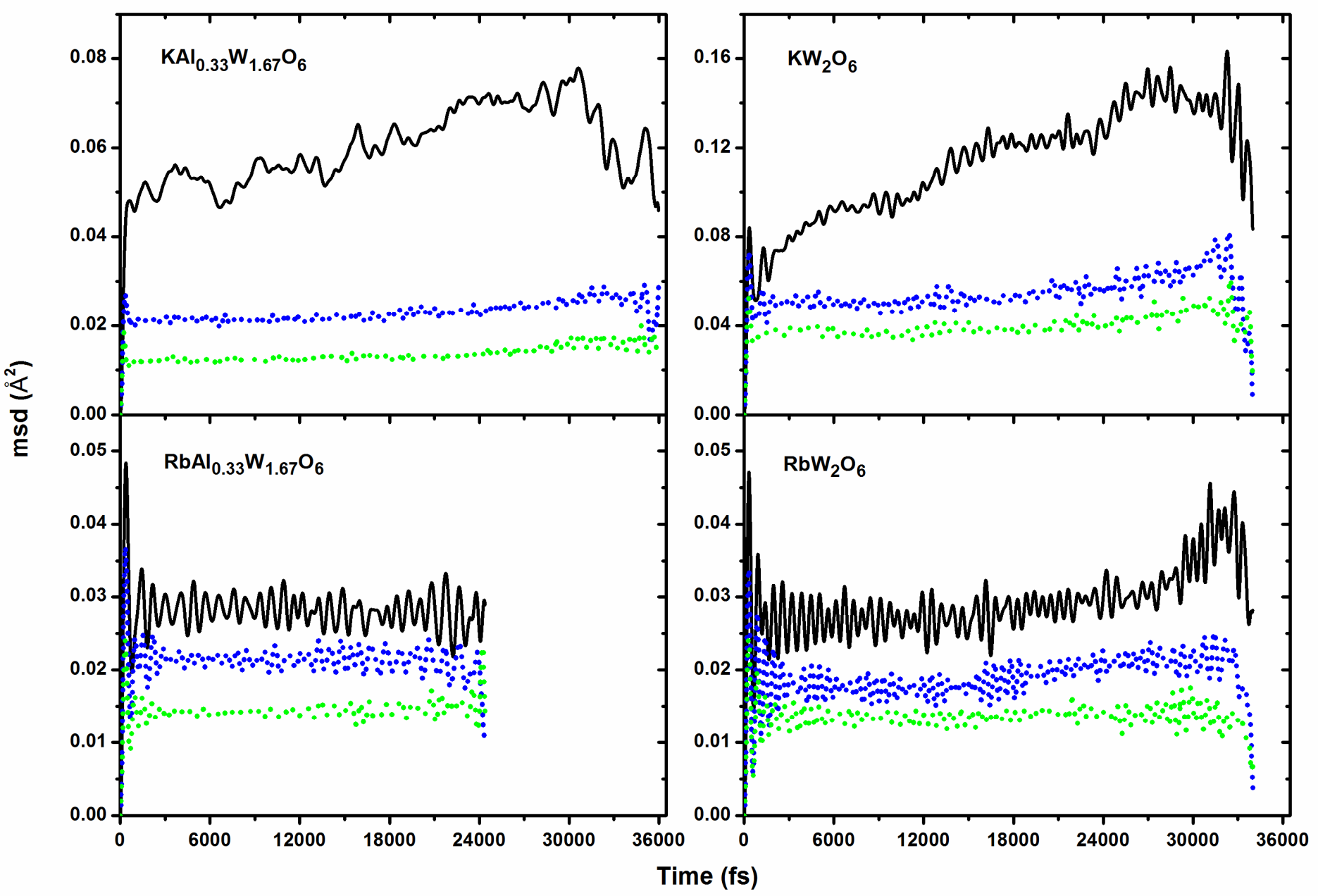

**FIG. 3. The msd autocorrelation functions, calculated from the MD simulations at 100 K in the principal axes coordinates, for K and Rb in the indicated compounds. For each compound, each curve corresponds to each of the three principal axis directions, with the top curve being the direction of maximum displacement. The K displacements are significantly larger than those of Rb, hence different scales are used to clearly highlight the features of each compound. A sharp distinction between the K and Rb dynamics is evident in the direction of maximum displacement where the K exhibits a diffusive profile consistent with local diffusion. Note that the simulation lengths are 36, 34, 24, and 34 ps for $KAl_{0.33}W_{1.67}$, $KW_2O_6$, $RbAl_{0.33}W_{1.67}O_6$, and $RbW_2O_6$, respectively.**

We now examine the consequences of the novel K rattling to the thermal conductivity of the Al-doped compounds. To be useful for thermoelectric materials development, these unusual K dynamics need to bring about a reduction in the thermal conductivity. The experimental thermal conductivity data are plotted in FIG.4 for the temperature range 5 – 300 K. As we show later, it is in this temperature range that the novel K dynamics involving local diffusion are most pronounced. The results in FIG. 4 indicate that the overall thermal conductivities of all three compounds are relatively low implying that they could be candidates for thermoelectric materials with further development. Considering the individual thermal conductivity curves, it is noted that the increase in thermal conductivity from Cs to Rb does not extend from Rb to K, with instead, the K conductivity being lower than that of Rb throughout the whole temperature range. There is a dip in the K-pyrochlore thermal conductivity at about 50 K giving it a concave upwards profile as it gradually approaches the Rb curve before the slope changes at T ~ 280 K.

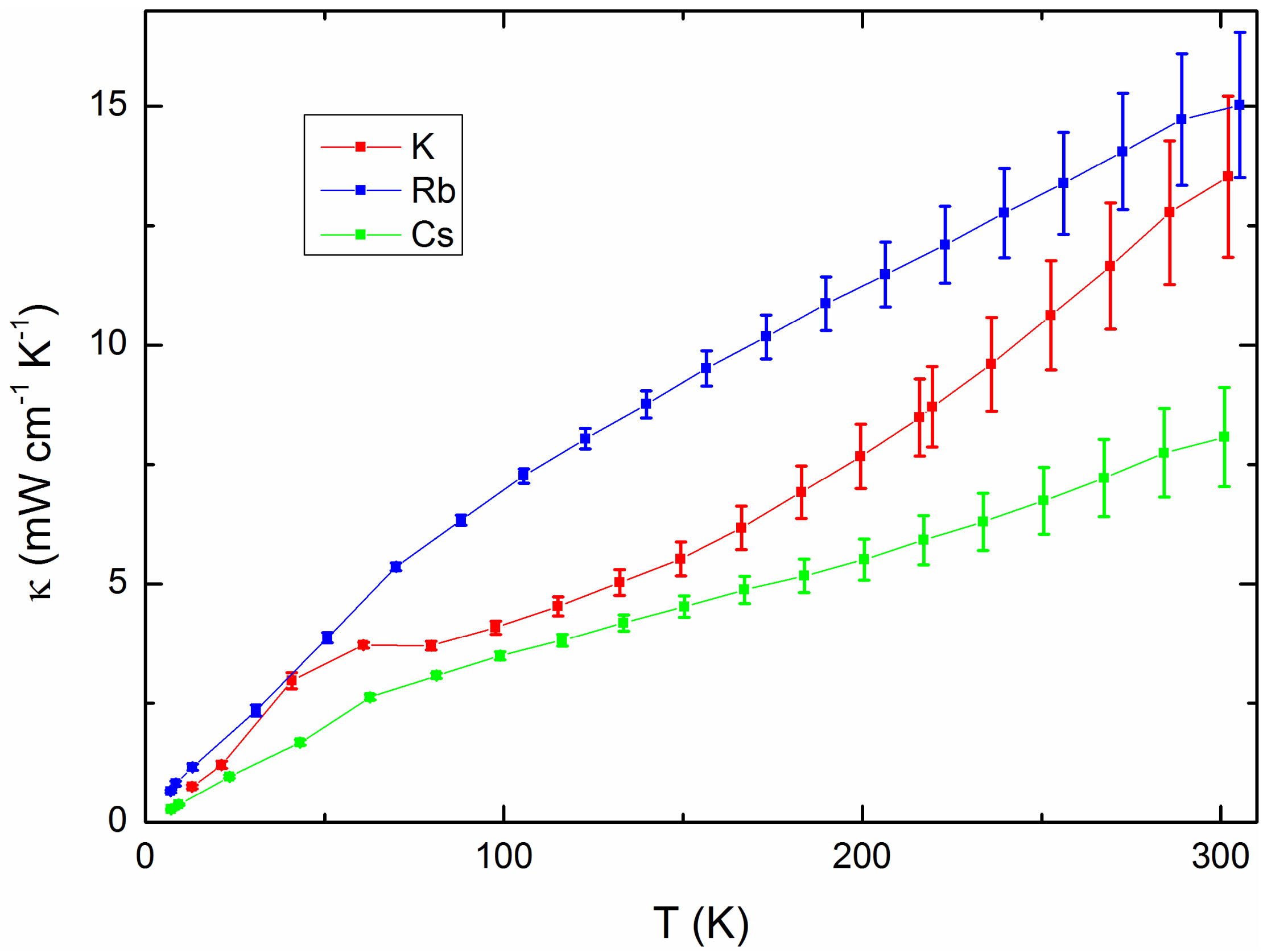

**FIG. 4 Thermal conductivity data for the Al-doped β-pyrochlore tungstates, $KAl_{0.33}W_{1.67}O_6$, $RbAl_{0.33}W_{1.67}O_6$, and $CsAl_{0.33}W_{1.67}O_6$ from 5 K to 300 K. Continuous lines are a guide to the eye only. The K thermal conductivity is much lower than would be expected from the increase exhibited in the conductivity from Cs to Rb. A more complex behaviour is also evident for the K conductivity, with a dip ~ 50 K < T < 250 K.**

Since, in general, the macroscopic thermal conductivity of a material is a sum of the electronic and lattice contributions, see Eq. (1), before discussing the role of the novel K dynamics in the unusual features of the K thermal conductivity curve, it is important to first show that the electronic contribution is negligible. Figure 5 shows the elemental contributions to the total density of states (TDOS) of the electrons where it is evident that all three compounds have similar electronic structures. Importantly, the large band gap of ~ 1 eV for all compounds implies that thermal excitations of electrons from the valence band into the conduction band are negligible at the temperatures of the thermal conductivity experiment. For this reason, in the following discussion, we neglect the electronic contribution to the thermal conductivity and attribute all features of the conductivity data to the lattice part.

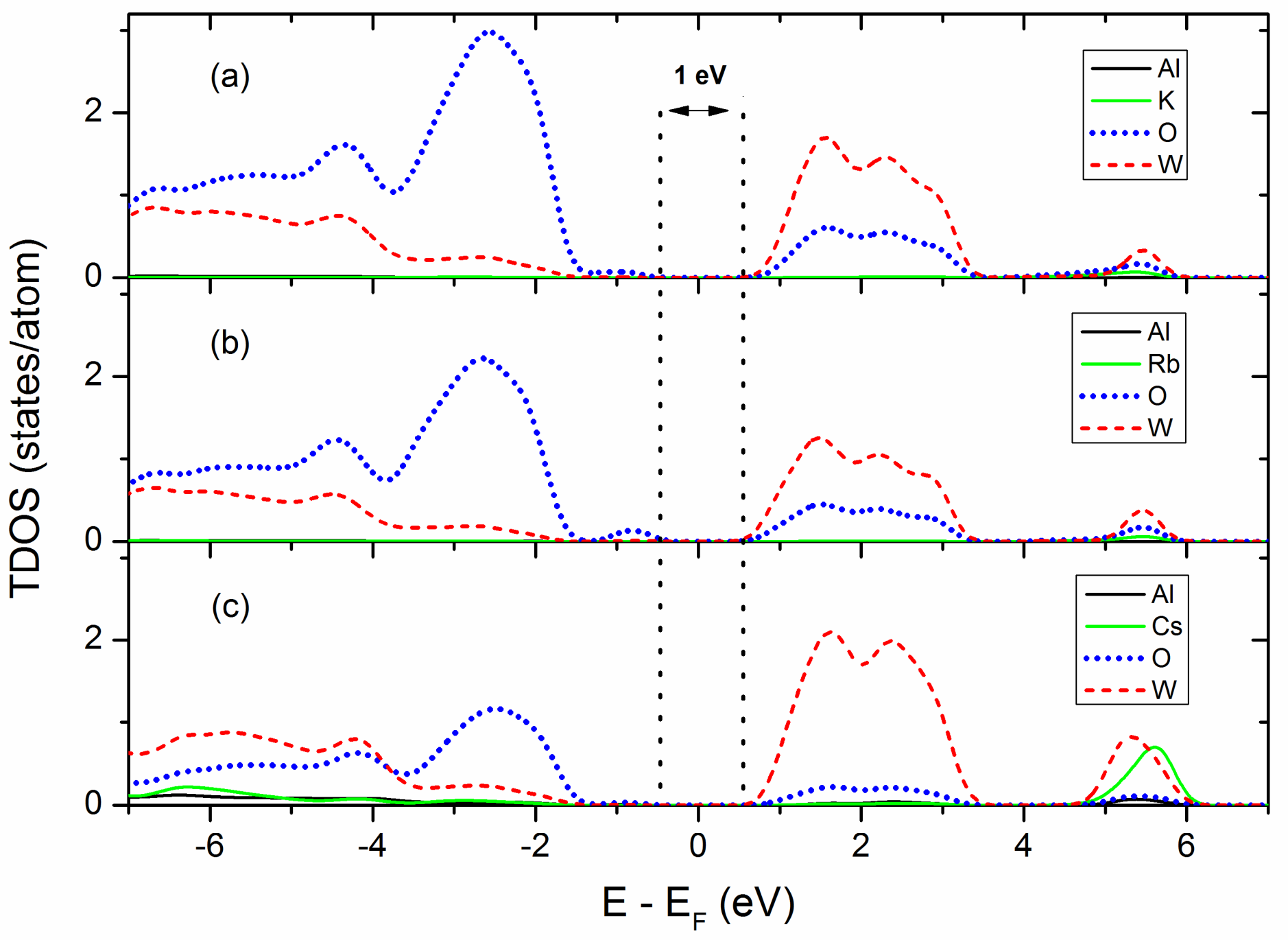

**FIG. 5. Electronic total (by element) density of states of (a) $KAl_{0.33}W_{1.67}O_6$, (b) $RbAl_{0.33}W_{1.67}O_6$, and (c) $CsAl_{0.33}W_{1.67}O_6$ near the Fermi level, where $E_F$ is the Fermi energy. All three compounds have similar electronic structures with a band gap of ~ 1 eV, indicating that electronic contributions to the thermal conductivity are negligible at low temperatures. The top of the valence band is predominantly O 2p states while the bottom of the conduction band is mainly the W 5d states.**

## *B. Novel Rattling K Dynamics and Thermal Conductivity*

Since we attribute the thermal conductivities of FIG. 4 to purely the lattice contribution, the differences across the K - Cs series should be related to the lattice dynamics. We showed previously[25] that the low-energy modes of the rattlers increase in frequency from Rb (with a double-peak feature centered at ~ 5 meV) to Cs centered at ~ 6 meV) but the situation for K is more complicated as the frequencies cover a wider range including frequencies below and above Rb and Cs, respectively. In comparing Cs and Rb, it is plausible that the slightly higher Cs modes are more effective at scattering the heat-carrying acoustic phonons of the lattice than the relatively low Rb modes. This argument suggests that the thermal conductivity of the K, with some even lower modes, should be higher than that of Rb. Nonetheless, the extension of the argument to the K case is complicated because the K vibrational density of states (VDOS) is spread over a much wider range. For the remainder of this discussion, we concentrate on explaining the interesting depression in the K thermal conductivity between T ~.50 K and T ~ 280 K in terms of the novel K dynamics.

In FIG. 6, we plot the temperature-dependent magnitude spectra of the K atoms in $KAl_{0.33}W_{1.67}O_6$ at six temperatures covering the range 25 – 400K as indicated. The 25 K spectrum exhibits oscillatory dynamics but at 50 K, a change to diffusive dynamics has occurred as evidenced by the emerging diffusive features in the spectrum at low energies. We cannot say the exact temperature at which the crossover from oscillatory to diffusive dynamics occurs but it seems to be close to 50 K. The diffusive dynamics become more pronounced with temperature as the relative magnitudes of the diffusive features increases but by 250 K, these features are weakening. By 400 K, a crossover back to oscillatory dynamics has occurred but again, we cannot state the exact temperature at which this change occurs.

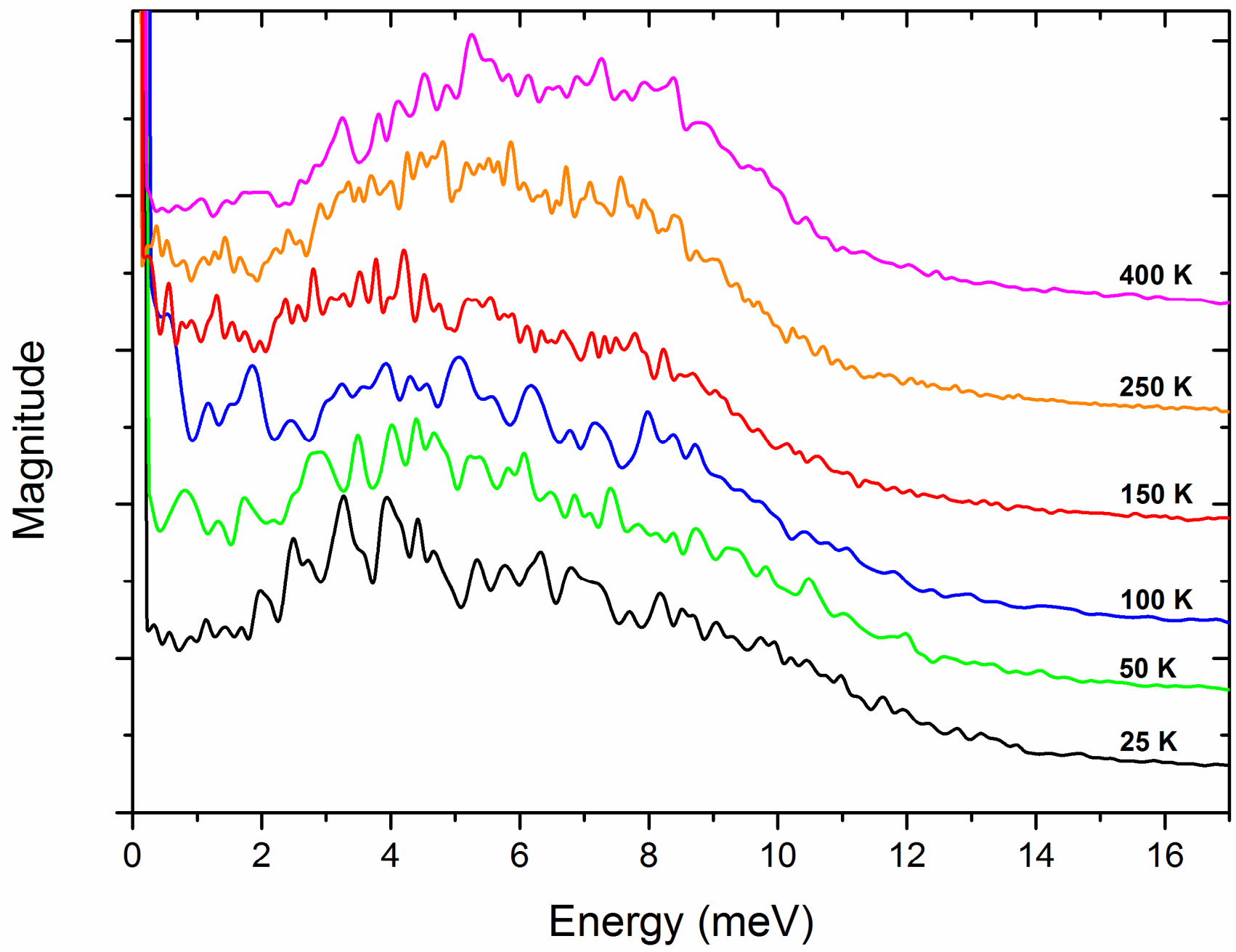

**Fig. 6. K magnitude spectra for $KAl_{0.33}W_{1.67}O_6$, calculated from the MD simulations at the different temperatures as indicated. The spectra exhibit a shift from oscillatory dynamics (25 K), to diffusive dynamics (50, 100, 150 K), and back to oscillatory dynamics (250, 400 K), mirroring the changes in the thermal conductivity in Figure 4.**

The shifts from oscillatory to diffusive and back to oscillatory dynamics exhibited in the magnitude spectra of FIG. 6 are seen more clearly in the corresponding msd functions plotted in FIG. 7. Of interest for discussing local diffusion is the profile of the msd functions in the direction of maximum displacement. Interestingly, at 25 K, maximum displacement occurs in two directions which can be understood from a consideration of the magnitudes of the displacements involved. The results suggest that at these relatively small displacements, the potential is degenerate in the two directions. However, the main point is that the profiles of the msd functions in both directions are relatively flat oscillating about some mean value consistent with oscillatory dynamics. At 50 K, the msd function for the direction of maximum displacement is now distinct although it merges with that for the second direction at ~ 24 ps, presumably as a result of the gradual cooling of the simulation. Nonetheless, the profile of the msd function for the direction of maximum displacement before the fall off does show the onset of local diffusion, from the relatively large rises and falls of the curve. By 100 K through to 150 K, the local diffusion in the direction of maximum displacement is significantly pronounced indicating that the atoms are diffusing over longer distances with increasing temperature as would be expected. However, somewhat surprisingly, by 250 K, the diffusive character of the msd function has virtually vanished and the K dynamics tend towards the harmonic limit. The profile at 400 K shows that the K dynamics increasingly become oscillatory with increasing temperature in this region. We examine this unexpected shift below but note here that the results show a strong connection between the depression in the thermal conductivity and local diffusion of the K atoms.

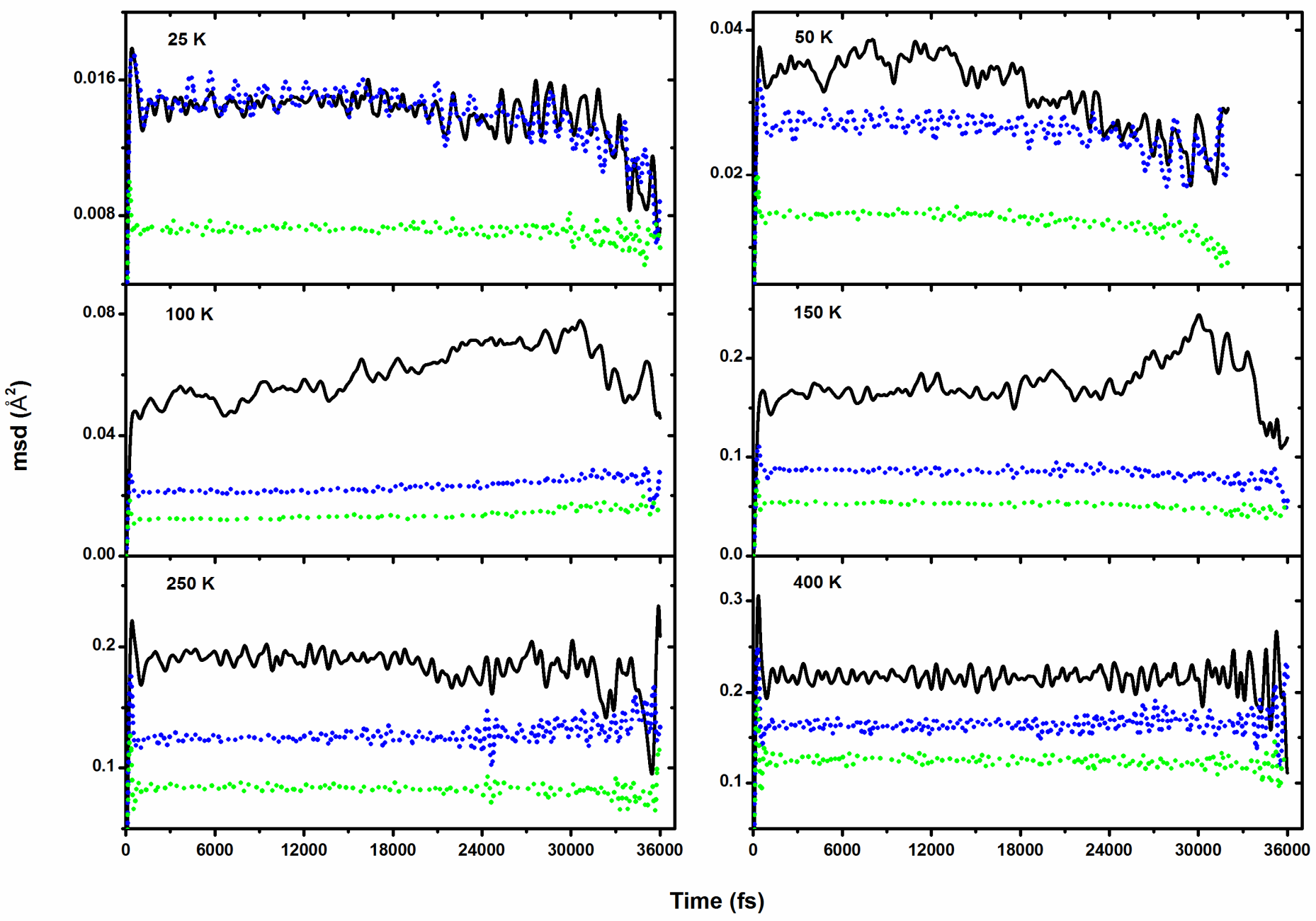

**FIG. 7. The K msd autocorrelation functions corresponding to the magnitude spectra in Figure 6, and calculated in the principal axes coordinates. For each compound, each curve corresponds to each of the three principal axis directions, with the top curve being the direction of maximum displacement. As the msd values are very different at the different temperatures, different scales are used. The shift from oscillatory dynamics (25 K) to local diffusion (50, 100, 150 K) and back to oscillatory dynamics (250, 400 K) is evident. The msd curves reveal that the local diffusion**

**occurs only in the direction of maximum displacement while the other directions always preserve the oscillatory dynamics. Part of the fall-off in the msd functions towards the end of the 25 K and 50 K simulations may be related to the cooling which would have a more significant impact on the displacements at these low temperatures compared to the higher temperatures. The 50 K simulation is 32 ps long while the rest are 36 ps.**

The dip in the K-pyrochlore thermal conductivity at about 50 K in FIG. 4 roughly coincides with the onset of the local diffusive dynamics of the K atoms as shown in FIG. 6 and 7. The slope of the K- pyrochlore thermal conductivity curve changes from concave upwards at ~ 280 K, roughly coinciding with the change to predominantly oscillatory dynamics of the K atoms. Thus, we conclude that the unusual decrease in the thermal conductivity associated with the concave upwards part of the K-pyrochlore curve is a consequence of the K atoms undergoing local diffusion. The absence of local diffusion at 25 K suggests that it is an activated process, and we estimate that the activation energy is in the order of ~ 50 K. The result that a rattler undergoing activated local diffusion in a suitable energy range can significantly reduce the lattice thermal conductivity of a material may be usefully exploited in the development of new materials for thermoelectric applications. It suggests that the goal of the multi-filler approach[23, 24] may be extended to include rattlers which reduce thermal conductivity through undergoing local diffusion.

Finally, we discuss the shift from diffusive to oscillatory dynamics at high temperatures exhibited in FIG. 6 and 7. To understand why local diffusion is unfavourable at higher temperatures, we examine the evolution with temperature of the local potential around the K atoms by calculating the PMFs at the various temperatures. The results are plotted in FIG. 8 where we also plot the average PMF of the lattice (i.e., Al, W, and O atoms) at the respective temperatures. The PMFs of the lattice are displaced by 0.61 Å so that the K and lattice PMFs just touch at 25 K. In order to highlight the features of the PMFs relevant to the issue of the diffusive/oscillatory dynamics crossover at high temperatures, we have restricted the plots to the energy range, 0 – 35 meV.

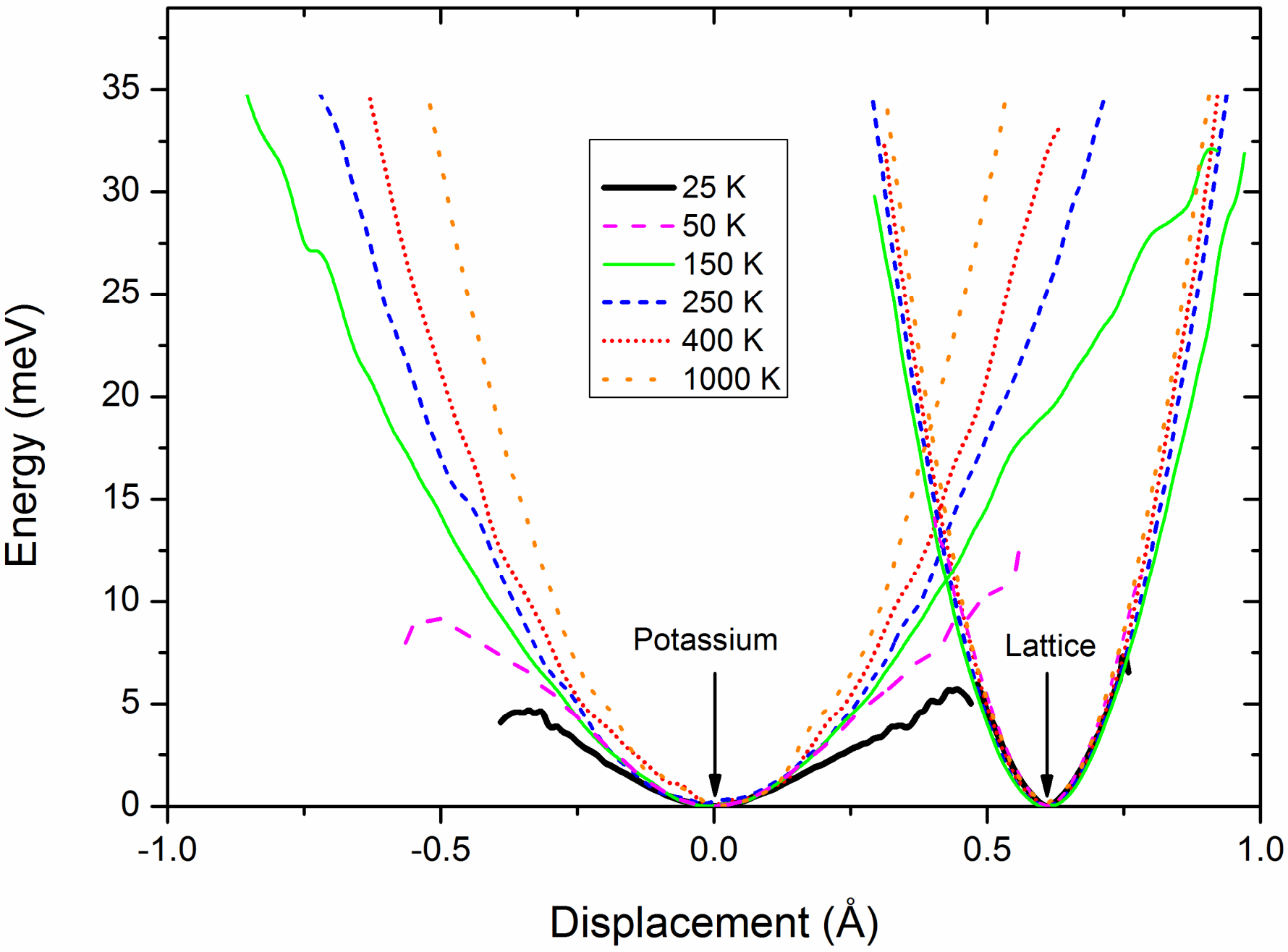

**FIG. 8. The K and lattice (Al, W and O) PMFs in $KAl_{0.33}W_{1.67}O_6$ calculated from the MD simulations in the direction of maximum displacement (for each atom) for the different temperatures as indicated. The lattice PMF is displaced by 0.61 Å so that both the lattice and K PMFs at 25 K just touch at 0.47 Å. As the temperature increases from 25 K, the lattice PMF expands into the region of the K PMF, forcing a narrowing of the latter which leads to a more complex temperature dependence of the K PMF.**

We note that FIG. 8 is an approximate schematic and a simple attempt to capture an average picture of the temperature dependence of the main dynamics of the system, viewed as composed of two subsystems, namely, the rattler (all the K atoms) and the cage (all Al, W, and O atoms). The simplified abstraction of the system represented in FIG. 8 implies that the cage framework is reduced to a single particle (an atom) experiencing a temperature-dependent local potential as shown. The crucial aspect of this potential is that it is primarily determined by the bonding within the cage framework and is virtually independent of the rattler because of the much weaker bonding to the latter. In FIG. 8, we represent this weak cage-rattler interaction by displacing the lattice PMF in such a way that the low-temperature (25 K) PMFs of the K and lattice just touch. Although we only show the positive displacement of the lattice PMF in FIG. 8, it is important to note that a similar, but negative, displacement is assumed in this discussion, so that the K PMF is sandwiched between the two lattice PMFs. Note also that in the principal axes basis used in this analysis, each atom will have its own direction of maximum displacement and these have all been aligned together for the calculation of the average PMFs and therefore FIG. 8 should not be interpreted to mean that the direction of maximum displacement for the rattler is the same as that of the lattice. In addition, the averaging of the PMFs over many atoms obscures some of the details in the single-site PMFs.[25] Nonetheless, this simple average picture reveals features of the PMFs which shed light on the unusual K diffusive/oscillatory dynamics crossover at high temperatures. We see that the K PMF at 25 K is shallow and highly anharmonic while the corresponding lattice PMF is significantly more harmonic. As the temperature increases, the behaviour of the K PMF is strikingly different from that of the lattice. Except for some minor discrepancies, the high temperature lattice PMFs are virtually extrapolations of the 25 K PMF to higher energies. In contrast, the K PMF virtually reconstructs at each temperature, exhibiting significant hardening with increasing temperature. This contrast in the behaviour of the PMFs reflects the fact that the lattice PMF is virtually independent of the rattler being controlled by the stronger bonding within the cage framework.

Thus the qualitative physical picture that emerges from this is that at low temperatures, the lattice PMF does not significantly extend out into the region of the K PMF permitting a softer shallower potential for the latter. It is conceivable that the K atoms could undergo diffusive dynamics in such a potential but this does not occur as their kinetic energy is too low. As the system temperature increases, the lattice PMF being harder and more harmonic than the K PMF extends further into the region of the latter forcing it inwards, i.e., hardening. By ~ 50 K, the K atoms have gained enough kinetic energy to explore a larger region while the lattice PMF has not yet extended out to restrict this motion. Consequently, the K atoms undergo local diffusion, a scenario which continues until ~ 250 K when the lattice PMF has eventually extended further into the region of the K PMF forcing enough hardening of the latter to prevent local diffusion. This simple qualitative picture suggests that there may be limitations in discussing finite-temperature dynamics based on potentials calculated at zero temperature for systems with PMFs that reconstruct rather than extrapolate from their low temperature values. This may be the case with the osmates analogue, $KOs_2O_6$, for which zero-temperature calculations have revealed a highly anharmonic potential with a flat bottom.[37-40]

# IV. Conclusion

In this work, we have shown, from *ab initio* MD that the novel K rattling dynamics in the defect tungstate pyrochlores play an important role in the lattice thermal conductivity of the material. In particular, we show that the local diffusion of the K atoms at intermediate temperatures leads to a significant reduction in the thermal conductivity. We suggest that this unusual mode of thermal conductivity reduction by a rattler may open a new route for the phonon engineering of cage compounds for thermoelectric applications by the multi-filler approach. We find two crossovers in the K rattling dynamics which are qualitatively explained by a complex temperature dependence of the potential around the K atoms.

**Acknowledgements**

We gratefully acknowledge helpful discussions with Dr R. Kutteh on various technical issues regarding VASP simulations.